\documentclass[a4paper]{article}
\usepackage[T1]{fontenc}
\usepackage{amsmath}

\let\oldvec=\vec
\def\vec#1{\oldvec#1}

\author{Éric Brunet\\\small\emph{Laboratoire de Physique Statistique, École
Normale Supérieure,}\\\small\emph{24 rue Lhomond, 75231 Paris Cedex 05, France}}
\title{Fluctuations of the winding number of a directed polymer in a
random medium}
\begin{document}
\maketitle
\begin{abstract}
For a directed polymer in a random medium lying on an infinite cylinder,
that is in $1+1$ dimensions with finite width and periodic boundary
conditions on the transverse direction, the winding number is simply the
algebraic number of turns the polymer does around the cylinder. This
paper presents exact expressions of the fluctuations of this winding
number due to, first, the thermal  noise of the system and, second, the
different realizations of the disorder in the medium.\\[1ex]
PACS numbers: 05.40.--a, 36.20.--r, 64.60.Cn, 71.55.Jv
\end{abstract}

\section*{Introduction}
A directed polymer in a random medium is one of the most simple
non-trivial disordered system and is, as such, of special theoretical
importance. Indeed, several exact results on directed polymers
with strong disorder have been obtained\cite{Kardar.87,
Halpin-HealyZhang.95, BrunetDerrida.00, BrunetDerrida2.00, 
EmigKardar.01}, so that general approximation schemes developed to tackle
more complicated disordered systems such as spin glasses could be tested
within the directed polymer context. The directed polymer is also
relevant in the context of nonequilibrium phenomena, as it is related,
through simple changes of variables, to growth models governed by the
Kardar-Parisi-Zhang (KPZ)
equation\cite{KardarParisiZhang.86, Krug.97} and to non-turbulent flows
such as the Astmmetric exclusion process
(ASEP) model\cite{Halpin-HealyZhang.95, Krug.97}.

The objective of this paper is to study the winding of a directed polymer
lying on the surface of a cylinder. The algebraic number of turns~$W$ the
polymer does around the cylinder is a random variable which depends on
the realization of the disorder and which fluctuates because of the
thermal noise. The statistics of the winding number of a polymer in an
homogeneous medium in~$2+1$ dimensions around a cylinder goes back to the
work of Spitzer\cite{Spitzer.58} and is relevant for the physics of
vortices in type~II superconductors\cite{DrosselKardar.96,
DrosselKardar.98, NelsonStern.97}. In physical situations, the system is,
however, usually
disordered; the effect of columnar defects has been studied
analytically\cite{NelsonStern.97}, and the winding number around a
cylinder of a polymer in a random medium with point-like disorder
in~$2+1$ dimension has been explored numerically\cite{DrosselKardar.96,
DrosselKardar.98}. When there is an attractive interaction between the
polymer and the winding center, the polymer can be confined around the
cylinder and the system can be regarded as a polymer in the~$1+1$ dimension
with periodic boundary conditions. In that situation, the present work
gives exact expressions for the statistics of the winding number.

The directed polymer on a
cylinder is also related to the classical limit of strongly correlated
fermions in one dimension with disorder (Luttinger liquids): the~$x$
position of the directed polymer corresponds to the phase of the
fermions, and the phase has, of course, periodic boundary conditions.
The winding number of the polymer corresponds to the density of fermions.
Disorder, while periodic in both cases, does not have exactly the same
correlations, but the models are sufficiently similar to hope for some
universality\cite{NattermannGiamarchiLeDoussal.03,GiamarchiOrignac.03}.

A first result of the present paper states that the thermal
fluctuations of the winding number are simply equal to what one would
obtain for a directed polymer in a homogeneous medium; disorder is
simply averaged out. A second result concerns the
thermal-averaged winding number~$\overline{W}$. Because of the randomness
of the medium, this quantity is not zero and the expression of its
variance~$\big\langle \overline{W}^2 \big\rangle$ averaged over disorder
is obtained.

The present paper is organized as follows: the first section is a brief
recall of how the directed polymer in a random medium can be mapped to a
quantum mechanical problem of interacting bosons using the replica method,
and how this quantum mechanical problem can be solved with the Bethe
Ansatz\cite{Kardar.87, BouchaudOrland.90, Halpin-HealyZhang.95,
BrunetDerrida.00, BrunetDerrida2.00}.
In a second section, the winding
number is introduced and defined  and the two main results of this paper
are stated in equations~(\ref{firstresult}) and~(\ref{secondresult2}).
Section three gives the main lines of the derivation, and, finally,
technical points are developed in the three appendixes.

\section{Definition, notations and free energy of a directed polymer}

Let us consider a directed polymer in $1+1$ dimensions where the
dimension in which the polymer is directed (the ``time'' dimension) is
taken to be very large and the transverse dimension (the ``space''
dimension) has width~$1$ and periodic boundary conditions. As it is
directed, the polymer can be described by a single-valued function~$y(t)$
and the partition function of a directed polymer of length~$t$ ending at
position~$x$ is given by
\begin{equation}
Z(x,t)=\int_{y(t)=x}\kern-2.0em{\cal
D}y(s)\,\exp\left(-\int_0^t\kern-0.7em ds\left[{1\over2}\left(dy\over
ds\right)^2+\eta\big( y(s),s\big)\right]\right),
\label{path}
\end{equation}
where~$\eta(x,t)$ is the contribution by the random medium to the energy
of the system. Disorder in the medium is assumed to be characterized
by an uncorrelated Gaussian noise of variance~$\gamma$:
\begin{align}
\big\langle \eta(x,t)\big\rangle &=0, &\big\langle
\eta(x,t)\eta(x',t')\big\rangle&= \gamma \delta(x-x')\delta(t-t'),
\label{variance}
\end{align} 
where the brackets $\langle\ \rangle$ represent the average over disorder.

It is a well known result\cite{Kardar.87, BouchaudOrland.90,
Halpin-HealyZhang.95, BrunetDerrida.00, BrunetDerrida2.00} that this
system can be successfully mapped to a quantum mechanical problem using
the replica method; indeed, if we define 
\begin{equation}
\psi(\vec x;t)=
	\psi(x_1,\dots,x_n;t)={\big\langle Z(x_1,t)\dots Z(x_n,t)\big\rangle
\over\big\langle Z(t)\big\rangle^n},
\label{psiZ}
\end{equation}
where 
\begin{equation}
Z(t)=\int \kern-0.3emdx\,Z(x,t)
\end{equation}
is the full partition function, then~(\ref{path}) and~(\ref{variance})
imply
\begin{equation}
{\partial\psi\over\partial
t}={1\over2}\sum_{i=1}^n{\partial^2\psi\over\partial
x_i^2}+\gamma\sum_{i<j}\delta(x_i-x_j)\psi,
\label{growth}
\end{equation}
with periodic boundary conditions on all the space variables~$x_i$:
\begin{equation}
\psi(\dots,x_i=0,\dots;t) = \psi(\dots,x_i=1,\dots;t).
\label{pbc}
\end{equation}
The normalization by~$\langle Z\rangle^n$
in~(\ref{psiZ}) is just a simple way to get rid of a low-scale divergence
introduced by the continuous description~(\ref{path}) of the system. In
other words, without this normalization, there would be a trivial extra
term in~(\ref{growth}) involving the lattice size of an underlying
discrete formulation of the problem.

For an infinitely long polymer, that is, in the large~$t$ limit,
the amplitude of $\psi(\vec x;t)$ is given by the fastest growing
mode of~(\ref{growth}, \ref{pbc}). In quantum mechanical language, we have
\begin{equation}
\begin{aligned}
\lim_{t\to\infty}{\log \psi(\vec x;t)\over t}&=
\lim_{t\to\infty}{\log \int dx_1\dots dx_n\,\psi(\vec x;t)\over t}\\
&=
\lim_{t\to\infty}{\log \big\langle Z(t)^n\big\rangle-n\log\big\langle
Z(t)\big\rangle\over t}\\
&= -E(n,\gamma),
\end{aligned}
\label{relates}
\end{equation}
where~$E(n,\gamma)$ is the ground-state energy of the Hamiltonian
\begin{equation}
{\cal H}=-{1\over2}\sum_{i=1}^n{\partial^2\over\partial
x_i^2}-\gamma\sum_{i<j}\delta(x_i-x_j),
\label{hamilton}
\end{equation}
which describes~$n$ particles with \emph{attractive} delta-interactions
on a ring of size~$1$.

The same Hamiltonian with a negative value of~$\gamma$ (that is, with a
\emph{repulsive} delta-interaction) has been much studied to determine
the spectrum of a gas of bosons\cite{LiebLiniger.63, YangYang.69,
Gaudin.71, Jimboetall.80, Thacker.81, Gaudin.83}. In that context, using
the Bethe Ansatz\cite{Bethe.31}, it was shown that the ground-state
energy of~(\ref{hamilton}) can be written as
\begin{equation}
E(n,\gamma)=-{1\over2}\sum_{\alpha=1}^n \lambda_\alpha^2,
\label{e1}
\end{equation}
where the~$\{\lambda_\alpha\}$ are solutions of
\begin{equation}
e^{\lambda_\alpha}=\prod_{\substack{1\le\beta\le
n\\\beta\ne\alpha}}{\lambda_\alpha-\lambda_\beta+\gamma
\over \lambda_\alpha-\lambda_\beta-\gamma},\qquad
\text{with}\qquad\lim_{\gamma\to0}\lambda_\alpha=0.
\label{e2}
\end{equation}
Of course, this expression obtained for~$\gamma<0$ remains
valid in the directed polymer context where~$\gamma>0$.

In the quantum mechanical problem~(\ref{hamilton}), the ground state
energy~$E(n,\gamma)$ is well defined only for integral~$n$; after all,
$n$ is the number of particles. However, for the directed polymer,
$\big\langle Z^n\big\rangle$, which is related to $E(n,\gamma)$
through~(\ref{relates}), can be defined for arbitrary values of~$n$. The
small-$n$ limit is of special importance here: as the directed
polymer is a disordered system, the free energy is a random variable
and~$\big\langle Z^n\big\rangle$ is the generating function of this free
energy. Indeed, we have
\begin{equation}
{\log\big\langle Z^n\big\rangle\over t}=n {\big\langle\log Z\big\rangle
\over t}
+{n^2\over2}{\big\langle\log^2 Z\big\rangle_c\over t}
+{n^3\over6}{\big\langle\log^3 Z\big\rangle_c\over t}+O(n^4),
\end{equation}
where $\big\langle\log Z\big\rangle/t$, $\big\langle\log^2
Z\big\rangle_c/t=\big(\big\langle\log^2
Z\big\rangle-\big\langle\log Z\big\rangle^2)/t$, etc. are the cumulants of
the free energy per unit-length of the directed polymer.
Thus, if we can generalize equations~(\ref{e1}, \ref{e2}) to arbitrary values
of~$n$, the expansion of~$E(n,\gamma)$ for small~$n$ gives,
using~(\ref{relates}), the distribution of the free energy of the directed
polymer\cite{EdwardsAnderson.75}.

This method was used\cite{Kardar.87} for the directed polymer on a space of
infinite width in the~$x$-direction. The Bethe Ansatz equations are then
much simpler than~(\ref{e2}) and one obtains\cite{McGuire.64,Kardar.87},
when~$n$ is an integer, $E(n,\gamma)=\gamma^2(n-n^3)/24$. This result was
used to argue that only the two cumulants $\big\langle\log
Z\big\rangle/t$ and $\big\langle \log^3 Z\big\rangle_c/t$ do not vanish
in the large~$t$ limit and that, therefore, the fluctuations of~$\log Z$
scale like $t^{1/3}$\cite{Kardar.85, HuseHenley.85, HuseHenleyFisher.85}.

When space has finite width, however, it is easy to see that
the free energy is an extensive function and that all its cumulants scale
like~$t$. In two previous papers\cite{BrunetDerrida.00,
BrunetDerrida2.00}, we solved the Bethe Ansatz equations~(\ref{e2}) and
computed the three first terms of the small~$n$ expansion
of~$E(n,\gamma)$. Up to the order~$n^2$, the result is
\begin{equation}
E(n,\gamma)=n\left({\gamma\over2}+{\gamma^2\over24}\right)
-{n^2\gamma^{3/2}\over4\sqrt{2}}\int_0^{+\infty}\kern-1.2emd\lambda\
{\lambda^2 e^{-\lambda^2/2}\over\tanh{\lambda\sqrt{\gamma}\over2\sqrt{2}}}+O(n^3),
\label{expE}
\end{equation}
so that, using~(\ref{relates}):
\begin{gather}
\lim_{t\to\infty}{\big\langle\log Z\big\rangle-
\log\big\langle Z\big\rangle\over
t}=-\left({\gamma\over2}+{\gamma^2\over24}\right),\\
\begin{aligned}
\lim_{t\to\infty}{\big\langle\log^2 Z\big\rangle_c\over
t}&={\gamma^{3/2}\over2\sqrt{2}}\int_0^{+\infty}\kern-1.2emd\lambda\
{\lambda^2
e^{-\lambda^2/2}\over\tanh{\lambda\sqrt{\gamma}\over2\sqrt{2}}}\\
&=\gamma+{\gamma^2\over12}-{\gamma^3\over360}+{\gamma^4\over5040}
	+O(\gamma^5)&&\text{for small~$\gamma$,}\\
&={\sqrt\pi\gamma^{3/2}\over4}+4\zeta(3)+O\left(1\over\gamma\right)
&&\text{for large~$\gamma$,}
\end{aligned}
\end{gather}
where~$\zeta(3)=\sum k^{-3}\approx1.20206$.

\section{Winding number of the directed polymer}

An important topological property of a directed polymer is its winding
number~$W$, that is the algebraic number of full turns the polymer makes
around the cylinder on which it lays.  One way to define this winding
number is to increase $W$ by one for each ``time''~$t$ where the~$x$
coordinate of the polymer goes from~$1^-$ to~$0^+$ and decrease $W$ by
one when~$x$ goes from~$0^+$ to~$1^-$. Another way is to unroll
the~$x$-coordinate and set~$W=\int\dot x\,dt$. Of course, the differences
between those two definitions smear out in the large~$t$ limit.

As for any quantity in a disordered system at finite temperature, the
winding number~$W$ fluctuates for two distinct reasons. One is the
thermal fluctuations: for a given realization of the disorder and at
finite temperature, the directed polymer fluctuates around the path with
the lowest energy, and those fluctuations may change the winding number
of the polymer. The other source of fluctuations is the quenched disorder
on the medium.

In this work, a horizontal bar is used to denote the thermal average, that
is the average  computed over all the possible directed polymers counted
with their Boltzmann weights.
The cumulants are noted with an extra~$c$ subscript:
\begin{align*}
\overline{W}&:\text{ thermal average of~$W$,}\\
\overline{(W^k)}_c&:\text{ $k$-th thermal cumulant of~$W$,}
\end{align*}
with $\overline{(W^2)}_c=\overline{W^2}- \overline{W}^2$,
     $\overline{(W^3)}_c=\overline{W^3}-3\overline{W}\,\overline{W^2}+
2\overline{W}^3$, etc.
These thermal averages and cumulants are calculated for a given, fixed,
realization of the disorder and usually depend on that realization.

The average and cumulants of a quantity~${\cal Q}$ computed over all the
realizations of the disorder are written with brackets:
\begin{align*}
\big\langle{\cal Q}\big\rangle&:\text{ average of~${\cal Q}$ computed over
all realizations of the disorder,}\\
\big\langle{\cal Q}^k\big\rangle_c&:\text{ $k$-th disorder cumulant of~$\cal Q$.}
\end{align*}

It is worth noting that, for a given realization of the disorder, the
thermal average~$\overline{W}$ of the winding number is \emph{not} zero;
the disorder breaks the symmetry and may favor one orientation over the
other. However, $\overline{W}$ is an extensive quantity and, if we
imagine that we cut an extremely long polymer in many very long sections,
all the sections are nearly independent and~$\overline{W}$ may be
regarded as the sum of uncorrelated random variables. Therefore,
\begin{equation}
\lim_{t\to\infty}{\overline{W}\over
t}=\lim_{t\to\infty}{\big\langle\overline{W}\big\rangle\over t}=0.
\end{equation}
This property that~$\overline W$ approaches~$\big\langle\overline
W\big\rangle$ in the large~$t$ limit is known as ``autoaveraging''.
Likewise, all the thermal cumulants of~$W$ (which are also extensive
quantities) share the same property:
\begin{equation}
\lim_{t\to\infty}{\overline{(W^k)}_c\over t}=
\lim_{t\to\infty}{\big\langle\overline{(W^k)}_c\big\rangle\over t},
\end{equation}
Those cumulants, which characterize the thermal fluctuations of a directed
polymer's winding number, depend on the realization of the disorder only
when the length~$t$ of the polymer is finite.

Other quantities of interest are the disorder cumulants of the
thermal-avera\-ge of the winding number of the polymer. Indeed, the
quantity~$\overline{W}$ depends on the realization of the disorder, and
its fluctuations are characterized by another series of cumulants:
\begin{equation}
\lim_{t\to\infty}{\big\langle\overline{W}^k\big\rangle_c\over t},
\end{equation}
with $\big\langle \overline{W}^2\big\rangle_c=\big\langle
\overline{W}^2\big\rangle-\big\langle
\overline{W}\big\rangle^2$, etc.
Actually, we might be interested in computing many quantities
characterizing the winding number, such as
\begin{equation}
\lim_{t\to\infty} {\big\langle
\overline{W^2}^2\big\rangle-\big\langle\overline{W^2}\big\rangle^2\over
t},
\end{equation}
which represents the fluctuations due to the disorder of the thermal-mean
square of the winding number, per unit length.

\bigbreak
A first result of the present paper is
\begin{equation}
\lim_{t\to\infty}{\big\langle\overline{(W^2)}_c\big\rangle\over t}=
1\qquad\text{and}\qquad
\lim_{t\to\infty}{\big\langle\overline{(W^k)}_c\big\rangle\over t}=
0\text{\ \ for $k\ne2$}.\label{firstresult}
\end{equation}
In other words, thermal fluctuations of the winding numbers are
Gaussian and independent of the disorder~$\gamma$. For an infinitely
long polymer, the thermal fluctuations of the winding number of the
polymer behave as if the directed polymer was simply doing a random walk
in a disorder-less environment.

\medbreak

A second result of the present paper is
\begin{equation}
\begin{aligned}
\lim_{t\to\infty}{\big\langle\overline{W}^2\big\rangle_c\over t}
&=\lim_{n\to0}\left[
{2\over n^2}\left({\partial
E(n,\gamma)\over\partial\gamma}-{2\over \gamma }
E(n,\gamma)\right)+{1\over n}-1\right],\\
&=-1+\left({2\over\gamma}-{\partial\over\partial\gamma}\right)
\lim_{t\to\infty}{\big\langle\log^2 Z\big\rangle_c\over t}.
\end{aligned}
\label{secondresult}
\end{equation}
where~$E(n,\gamma)$ is the ground state energy of the quantum problem
computed in~\cite{BrunetDerrida.00, BrunetDerrida2.00} and
given in~(\ref{expE}). Therefore:
\begin{equation}
\begin{aligned}
\lim_{t\to\infty}{\big\langle\overline{W}^2\big\rangle_c\over t}
&={\sqrt{\gamma}\over\sqrt{2}}\left(\int_0^{+\infty}\kern-1.7emd\lambda\
{\lambda^2 e^{-\lambda^2/2}\over\tanh{\lambda\sqrt{\gamma}\over2\sqrt{2}}}
-{1\over4}\int_0^{+\infty}\kern-1.7emd\lambda\
{\lambda^4 e^{-\lambda^2/2}\over\tanh{\lambda\sqrt{\gamma}\over2\sqrt{2}}}
\right)-1,\\
&={\sqrt{\pi\gamma}\over8}-1+{8\zeta(3)\over\gamma}+O(\gamma^{-2})
\qquad\text{for large $\gamma$},\\
&=\frac{{\gamma }^2}{360} - \frac{{\gamma }^3}{2520} + 
  \frac{{\gamma }^4}{16800} +O(\gamma^5)
\qquad\text{for small $\gamma$}.
\end{aligned}
\label{secondresult2}
\end{equation}
(where $\zeta(3)=\sum k^{-3}\approx 1.20206$.)

The expression~(\ref{path}) of the directed polymer's free energy is
written with dimensionless variables. If we explicitly put back physical
constants and use the following expression instead of~(\ref{path}):
\begin{equation}
Z(x,t)=\int_{y(t)=x}\kern-2.0em{\cal
D}y(s)\,\exp\left(-\beta\int_0^t\kern-0.7em ds\left[{\kappa\over2}\left(dy\over
ds\right)^2+\eta\big( y(s),s\big)\right]\right),
\end{equation}
where~$\beta=(k_BT)^{-1}$ is the inverse of temperature,~$\kappa$ is the
rigidity modulus of the line and where the spatial dimension~$x$ has
finite width~$w$ and periodic boundary conditions,
then~(\ref{firstresult}) and~(\ref{secondresult2}) become:
\begin{equation}
\begin{gathered}
\lim_{t\to\infty}{\big\langle\overline{(W^2)}_c\big\rangle\over t}=
{1\over\beta\kappa w^2}\qquad\text{and}\qquad
\lim_{t\to\infty}{\big\langle\overline{(W^k)}_c\big\rangle\over t}=
0\text{\ \ for $k\ne2$},\\
\lim_{t\to\infty}{\big\langle\overline{W}^2\big\rangle_c\over t}
={1\over\beta\kappa w^2}F(\beta^3\kappa w\gamma),
\end{gathered}
\end{equation}
where~$F(\gamma)$ is the scaling function given in~(\ref{secondresult2}).
We obtain the following expansions:
\begin{equation}
\begin{aligned}
\lim_{t\to\infty}{\big\langle\overline{W}^2\big\rangle_c\over t}
&\approx{\sqrt{\pi\beta\gamma}\over8\sqrt{\kappa} w^{3/2}}&&\text{at low
temperature,}\\
&\approx {\beta^5 \kappa\gamma^2\over360}&&\text{at high temperature.}
\end{aligned}
\end{equation}

\section{Derivation of (\ref{firstresult}) and (\ref{secondresult})}

\subsection{Equivalence to a quantum mechanical problem}

To obtain both results~(\ref{firstresult}) and~(\ref{secondresult}), 
we define a new partition function~$Z_z(x,t)$, the purpose of which is to
count the winding number of the polymer:
\begin{equation}
Z_z(x,t)=\int_{y(t)=x}\kern-2.0em{\cal D}y(s)\,
	e^{-\Big(\text{energy of path $y(s)$}\Big)
	+z\Big(\text{winding number of that path}\Big)}.
\label{defZz}
\end{equation}
The sum is made over all the directed polymers ending in~$x$, and the
``energy of a path'' is the same as in~(\ref{path}).

Clearly, $Z_z(t)=\int Z_z(x,t)\,dx$ is related to the winding number~$W$ by
\begin{equation}
Z_z(t)=Z_0(t)\overline{e^{zW}}.
\label{ZzW}
\end{equation}

If we define the winding number~$W$ as an integer that changes by
$\pm1$ each time the directed polymer wraps around the domain by
crossing the~$x=0$ or~$x=1$ boundary, then the
boundary conditions for~$Z_z$ is:
\begin{equation}
Z_z(0,t) = e^z Z_z(1,t).
\end{equation}
Apart from that, the equations satisfied by~$Z_z(x,t)$ are the same as the
equations satisfied by~$Z(x,t)$. In particular, if we define
\begin{equation}
\psi_{z_1,\dots,z_n}(x_1,\dots,x_n;t)
={\big\langle Z_{z_1}(x_1,t)\dots Z_{z_n}(x_n,t)\big\rangle
\over \big\langle Z_0(t)\big\rangle^n},
\end{equation}
this new wavefunction~$\psi$ is also a solution
of~(\ref{growth}); only the boundary conditions are changed: instead
of~(\ref{pbc}), the new conditions read:
\begin{equation}
\psi_{z_1,\dots,z_n}(x_1,\dots, x_i=0,\dots,x_n;t)=
e^{z_i}\psi_{z_1,\dots,z_n}(x_1,\dots, x_i=1,\dots,x_n;t).
\label{nbc}
\end{equation}
Thus, as in~(\ref{relates}), the long ``time'' $t$ behavior of~$Z_z(t)$
is given by
\begin{equation}
\lim_{t\to\infty}{\log\big\langle Z_{z_1}\dots
Z_{z_n}\big\rangle-n\log\big\langle Z_0\big\rangle\over
t}=-E(n,\gamma;z_1,\dots, z_n),
\label{nrelate}
\end{equation}
where~$E(n,\gamma;z_1,\dots, z_n)$ is the ground state energy of the same
Hamiltonian~(\ref{hamilton}) as before, but with the new boundary
conditions~(\ref{nbc}).

This new ground state energy~$E$ contains all the
information on  the winding number~$W$. For instance, from~(\ref{ZzW}),
and by definition of the cumulants, we have, for~$k>0$,
\begin{equation}
\overline{(W^k)}_c=\left.{\partial^k\over\partial z^k}\log
Z_z(t)\right|_{z=0}.
\end{equation}
$\big\langle\log Z_z\big\rangle$ is easily obtained from~(\ref{nrelate}):
we set all the~$\{z_i\}$ to one single value~$z$, make a small~$n$
expansion, and retain only the first order. We get:
\begin{equation}
\lim_{t\to\infty}{\big\langle\overline{(W^k)}_c\big\rangle\over t}
=-\lim_{n\to0}\left.{\partial^k\over\partial z^k}{\partial\over\partial n}
E(n,\gamma;z,\dots,z)\right|_{z=0}.
\label{firstresultE}
\end{equation}

Getting~$\overline{W}^2$ is more tricky. We would need~$(\partial \log
Z_z /\partial z)^2$, but that quantity can only be obtained
from~(\ref{nrelate}) if the parameters~$\{z_i\}$ take at least two
different values. For example, we have
\begin{equation}
\begin{aligned}
\lim_{n\to0}\big\langle Z_{z_1}Z_z^{n-1}\big\rangle
	&=\left\langle Z_0\Big(1+z_1\overline{W}+O(z_1^2)\Big)
	        \over  Z_0\Big(1+z  \overline{W}+O(z^2)\Big)\right\rangle,\\
	&= 1+(z_1-z)\big\langle\overline{W}\big\rangle
		-z
z_1\big\langle\overline{W}^2\big\rangle+O(z_1^2)+O(z^2),
\end{aligned}
\end{equation}
and
\begin{equation}
\lim_{n\to0}\log\big\langle Z_{z_1}Z_z^{n-1}\big\rangle
	= (z_1-z)\big\langle\overline{W}\big\rangle
	   -z z_1 \Big(\big\langle\overline{W}^2\big\rangle
			-\big\langle\overline{W}\big\rangle^2\Big)
+O(z_1^2)+O(z^2).
\end{equation}
Therefore, putting all the pieces together,
\begin{equation}
\lim_{t\to\infty}{\big\langle\overline{W}^2\big\rangle
                        -\big\langle\overline{W}\big\rangle^2\over t}
= \lim_{n\to0}\left.{\partial^2\over\partial z\partial z_1}
	E(n,\gamma;z_1,z,\dots,z)\right|_{\substack{z=0\\z_1=0}}.
\label{secondresultE}
\end{equation}

Finally, to obtain the results announced, we need to
compute~$E(n,\gamma;z,\dots,z)$ and, to the first order in~$z$ and~$z_1$,
$E(n,\gamma;z_1,z,\dots,z)$.

\subsection{Determination of $E(n,\gamma;z,\dots,z)$.}

\label{firstpart}
When all the parameters~$\{z_i\}$ are equal to one single value~$z$, the
problem is easy: all the replica play a symmetric role, so that the
ground state eigenvector~$\psi_{z,\dots,z}(x_1,\dots,x_n)$ of the
Hamiltonian~(\ref{hamilton}) is a symmetric function of all
the~$\{x_i\}$. As shown in appendix~\ref{zzz}, the standard Bethe Ansatz
derivation gives the result. Instead of~(\ref{e1}, \ref{e2}), we get
\begin{equation}
E(n,\gamma;z,\dots,z)=-{1\over2}\sum_{\alpha=1}^n\lambda_\alpha^2
\label{defEzzz}
\end{equation}
where the~$\{\lambda_\alpha\}$ are solutions of
\begin{equation}
e^{\lambda_\alpha+z}=\prod_{\substack{1\le\beta\le
n\\\beta\ne\alpha}}{\lambda_\alpha-\lambda_\beta+\gamma
\over
\lambda_\alpha-\lambda_\beta-\gamma}\qquad
\text{with}\qquad\lim_{\substack{\gamma\to0\\z\to0}}\lambda_\alpha=0.
\label{ba2}
\end{equation}
If we define
\begin{equation}
\tilde\lambda_\alpha=\lambda_\alpha  + z,
\end{equation}
then the~$\{\tilde\lambda_\alpha\}$ are clearly solutions of the standard
Bethe Ansatz equations~(\ref{e2}). Using~(\ref{defEzzz}), we obtain
\begin{equation}
E(n,\gamma;z,\dots,z)=E(n,\gamma)-{n\over2}z^2.
\end{equation}
$E(n,\gamma)=E(n,\gamma;0,\dots,0)$ is the ground state
energy~(\ref{expE}) before introduction of the~$\{z_i\}$. We have used
$\sum\tilde\lambda_\alpha=0$, which can be easily
deduced\cite{BrunetDerrida.00, BrunetDerrida2.00} from~(\ref{e2}).

Using~(\ref{firstresultE}), the result~(\ref{firstresult}) on the
thermal cumulants of the winding number is then immediate.

This method, based on a Bethe Ansatz, is not the simplest way to
obtain~(\ref{firstresult}). Indeed, the result could be obtained using
the \emph{statistical tilt symmetry}\cite{SchulzVillainBrezinOrland.88,
HwaFisher.94} of the problem; we define the winding number~$W$ of a
path~$y(s)$ as being simply the unrolled coordinate:
\begin{equation}
W=\int_0^t ds\ {dy\over ds}.
\end{equation}
(This new definition is, of course, equivalent to the previous one in the
large~$t$ limit.)
The change of variable~$y(s)=\tilde y(s)+zs$ in the
definition~(\ref{defZz}) of~$Z_z$ gives then:
\begin{equation}
Z_z(t)= e^{z^2\over2}\int{\cal
D}\tilde y(s)\,\exp\left(-\int_0^t\kern-0.7em
ds\left[{1\over2}\left(d\tilde y\over
ds\right)^2+\eta\big( \tilde y(s)+zs,s\big)\right]\right).
\end{equation}
Clearly~$\eta\big( \tilde y(s)+zs,s\big)$ have the same statistical
properties of~$\eta\big( \tilde y(s),s\big)$ and one gets
\begin{equation}
\big\langle\log Z_z(t)\big\rangle={z^2\over2}+\big\langle\log
Z_0(t)\big\rangle,
\end{equation}
from which the result~(\ref{firstresult}) is straightforward. The first
derivation with the Bethe Ansatz was included here as it demonstrates
part of the method used to obtain the second result~(\ref{secondresult}),
which cannot be derived from a \emph{statistical tilt symmetry} argument

\subsection{Determination of~$E(n,\gamma;z_1,z,\dots,z)$.}

When the parameters~$\{z_i\}$ are not identical, the wave
function~$\psi$ is no longer a symmetric function of the~$\{x_i\}$
and the problem is much more complicated. Therefore, the standard
bosonic Bethe Ansatz used in the previous case will not work. However, as
shown in appendix~\ref{z1zz}, using a more general Bethe Ansatz
that was first introduced to deal with non-bosonic
particles\cite{Yang.67, Sutherland.68, Takahashi.70}, we get the
following result:
\begin{equation}
E(n,\gamma;z_1,z,\dots,z)=-{1\over2}\sum_{\alpha=1}^n\lambda_\alpha^2,
\label{Ez1zzlambda}
\end{equation}
where the~$\{\lambda_\alpha\}$ are solutions of
\begin{equation}
e^{\lambda_\alpha+\zeta_\alpha}=\prod_{\substack{1\le\beta\le
n\\\beta\ne\alpha}}{\lambda_\alpha-\lambda_\beta+\gamma
\over
\lambda_\alpha-\lambda_\beta-\gamma}\qquad
\text{with}\qquad\lim_{\substack{\gamma\to0\\z\to0}}\lambda_\alpha=0,
\label{nba}
\end{equation}
and where the $\{\zeta_\alpha\}$ are such that
\begin{equation}
\begin{gathered}
e^{z}\prod_{\beta=1}^n\left[e^{\zeta_\alpha}
		+{\lambda_\alpha-\lambda_\beta\over\gamma}
			\left(e^z-e^{\zeta_\alpha}\right)\right]
= e^{z_1}\prod_{\beta=1}^n\left[e^{z}
		+{\lambda_\alpha-\lambda_\beta\over\gamma}
			\left(e^{z}-e^{\zeta_\alpha}\right)\right],\\
\lim_{\{z_i\}\to0}\zeta_\alpha=0.
\end{gathered}
\label{2prod}
\end{equation}
When~$z_1=z$, we recover~$\zeta_\alpha=z$, the result of the previous
section.

To determine the fluctuation~$\big\langle\overline{W}^2\big\rangle$ of the
winding number of the polymer, we only need to
compute~$E(n,\gamma;z_1,z,\dots,z)$ to the second order in the~$\{z_i\}$.
From~(\ref{2prod}), we easily get
\begin{equation}
\zeta_\alpha={z_1+(n-1)z\over
n}-{n\lambda_\alpha-\sum_{k=1}^n\lambda_k\over\gamma
n^3}(z_1-z)^2+O(\{z_i\}^3).
\end{equation}
We define, for all~$\alpha$,
\begin{equation}
\begin{aligned}
\tilde\lambda_\alpha&=\lambda_\alpha+\zeta_\alpha,\\
&=\lambda_\alpha\left(1-{(z_1-z)^2\over\gamma n^2}\right)+
{z_1+(n-1)z\over
n}+{\sum_{k=1}^n\lambda_k\over\gamma n^3}(z_1-z)^2,
\end{aligned}
\label{lambdatilde}
\end{equation}
and
\begin{equation}
\tilde \gamma=\gamma\left(1-{(z_1-z)^2\over\gamma n^2}\right).
\label{gammatilde}
\end{equation}
Using those new variables into~(\ref{nba}), we obtain the familiar
Bethe Ansatz equations:
\begin{equation}
e^{\tilde\lambda_\alpha}=\prod_{\substack{1\le\beta\le n\\\beta\ne\alpha}}
{\tilde\lambda_\alpha-\tilde\lambda_\beta
		+\tilde\gamma \over
\tilde\lambda_\alpha-\tilde\lambda_\beta-\tilde\gamma}+O(\{z_i\}^3),
\label{ba3}
\end{equation}
so that\cite{BrunetDerrida.00, BrunetDerrida2.00}, using the ground
state energy~$E(n,\gamma)$ given by~(\ref{expE}),
\begin{equation}
\sum_{\alpha=1}^n{\tilde\lambda_\alpha}^2=-2E(n,\tilde\gamma)+O(\{z_i\}^3)
\quad\text{and}\quad\sum_{\alpha=1}^n\tilde\lambda_\alpha=O(\{z_i\}^3).
\end{equation}

From there, using~(\ref{lambdatilde}), 
one can write~$\sum\lambda_\alpha^2$.
We finally get
\begin{equation}
\begin{aligned}
E(n,\gamma;z_1,z,\dots,z)
&=E(n,\gamma)-{1\over2} {\big[z_1+(n-1)z\big]^2\over n}\\
&\hphantom{=}+{1\over
n^2}\left[{2\over\gamma}E(n,\gamma)-{\partial
E(n,\gamma)\over\partial\gamma}\right](z_1-z)^2+O(\{z_i\}^3).
\end{aligned}
\label{Ez1zz}
\end{equation}
Then, finally, from~(\ref{secondresultE}), we get the announced
result~(\ref{secondresult}).

\section*{Conclusion}
Using the replica method with the directed polymer, one obtains a bosonic
quantum mechanical problem which can be solved by the Bethe Ansatz.
By extending this method and using a more general Bethe Ansatz that was
introduced to deal with non-bosonic particles\cite{Yang.67}, it has been
shown how the different quantities characterizing the fluctuations of the
directed polymer's winding number can be computed using new Bethe Ansatz
equations. Building upon a previous work\cite{BrunetDerrida.00,
BrunetDerrida2.00}, those equations were explicitly solved in
two cases giving the results~(\ref{firstresult})
and~(\ref{secondresult}--\ref{secondresult2}). The second result is
particularly interesting as it simply relates through
equation~(\ref{secondresult}) the fluctuations of the thermal-averaged
winding number and the fluctuations of the free energy of the directed
polymer. It would be interesting to understand this relation in a more
direct way.

In principle, the method presented in the present paper should allow to
compute more cumulants of the winding number and, eventually,
its complete probability distribution. For that, however, one needs, as a
first step, to generalize~(\ref{Ez1zz}) and
write the expansion of~$E(n,\gamma;z_1,z,\dots,z)$ to higher orders in
the~$\{z_i\}$. Indeed, one can show that, for example,
\begin{equation}
\lim_{t\to\infty}{\big\langle\overline{W^3}\ 
	\overline{W^{\vphantom3}}\big\rangle
-3\big\langle\overline{W^2}\big\rangle\big\langle\overline{W}^2\big\rangle
\over t}
= \lim_{n\to0}\left.{\partial^4\over\partial z\partial z_1^3}
        E(n,\gamma;z_1,z,\dots,z)\right|_{\substack{z=0\\z_1=0}}.
\end{equation}
Obtaining~$E(n,\gamma;z_1,z,\dots,z)$ to the fourth order in
the~$\{z_i\}$ is not, however, an easy task,
as the trick used in~(\ref{lambdatilde}) would not work at that order.

As a second step, to compute more complicated cumulants of the winding
number such
as~$\big\langle\overline{W}^4\big\rangle-3\big\langle\overline{W}^2\big\rangle^2$,
one needs to generalize~(\ref{Ez1zzlambda}--\ref{2prod}) to the case
where the~$\{z_i\}$ take at least four different values. Higher order
cumulants would require, of course, the energy of the system with
more different values of the~$\{z_i\}$. A matrix approach such as the one
developed in the present paper could lead to the result. Another
possibility might be to try an approach similar to the ``Nested
Bethe Ansatz'' method developed by Yang and Sutherland\cite{Yang.67,
Sutherland.68, Takahashi.70, Kardar.87, EmigKardar.01} to compute the
ground-state energy of the system described by the
Hamiltonian~(\ref{hamilton}) with different types of particles and
symmetry relations which depend on the type of particles. Their results
are not directly applicable to the directed polymer's winding number as
all the particles have the same symmetry relations but different
boundary conditions, but it might be worth investigating if the nested
Bethe Ansatz could be adapted.

\section*{Acknowledgments}
I would like to thank Pierre Le Doussal, who suggested this work, and
Bernard Derrida for his interesting discussions.

\appendix 
\makeatletter\@addtoreset{equation}{section}\makeatother
\def\theequation{\Alph{section}\arabic{equation}}

\section{Bethe Ansatz equations when all the~$z_i$ have the same
value~$z$} \label{zzz}

When all the~$\{z_i\}$ are equal to~$z$, all the particles
have symmetric roles and the standard bosonic Bethe Ansatz leads to
the result. To recall the standard
derivation\cite{LiebLiniger.63,Gaudin.71}, we look for solutions of
the following form:
\begin{equation}
\psi_{z,\dots,z}(x_1,\dots,x_n)
=\sum_{\sigma}a(\sigma)e^{\sum_{i=1}^n\lambda_{\sigma(i)}x_{\tau(i)}},
\label{wavefunction}
\end{equation}
where the sum is made over the~$n!$ permutations~$\sigma$
of~$\{1,\dots,n\}$ and where~$\tau$ is the permutation defined by
\begin{equation}
x_{\tau(1)} < x_{\tau(2)} < \dots < x_{\tau(n)}.
\label{deftau}
\end{equation}
The $n!$ amplitudes~$\{a(\sigma)\}$ and the~$n$
pseudo-wave-numbers~$\{\lambda_\alpha\}$ are unknown variables to be
determined.

We use this expression of~$\psi$ in~${\cal H}\psi=E\psi$, where~$\cal H$
is the Hamiltonian~(\ref{hamilton}). In the regions where all
the~$\{x_i\}$ are different, it is straightforward to get
\begin{equation}
E(n,\gamma;z,\dots,z)=-{1\over2}\sum_{\alpha=1}^n\lambda_\alpha^2,
\label{defE}
\end{equation}
so that we only need to determine the~$\{\lambda_\alpha\}$.
At each crossing of two particles, we have to ensure the correct
discontinuities in the derivatives of~$\psi$ to compensate for
the~$\delta$-functions in~$\cal H$. This gives the following conditions,
for all~$\sigma$ and all~$1\le k<n$:
\begin{equation}
a(\sigma\circ T_k)={\lambda_{\sigma(k)}-\lambda_{\sigma(k+1)}-\gamma
	\over	    \lambda_{\sigma(k)}-\lambda_{\sigma(k+1)}+\gamma}
a(\sigma),
\label{rela1}
\end{equation}
where~$T_k$ is the permutation that swaps~$k$ and~$k+1$ and leaves all
the other integers unchanged.

As any permutation~$\sigma$ can be written as a product of the elementary
permutations~$T_k$, one can use~(\ref{rela1}) to write all
the~$\{a(\sigma)\}$ up to an arbitrary multiplicative factor. However, as
the decomposition of a permutation $\sigma$ as a product of~$T_k$ is not
unique, one must check that the~$(n-1)n!$ equations~(\ref{rela1}) are
self consistent. The best way to do that is to write down explicitly
the solution:
\begin{equation}
a(\sigma)=\prod_{1\le \alpha<\beta\le
n}{\lambda_{\sigma(\alpha)}-\lambda_{\sigma(\beta)}
+\gamma\over\lambda_{\sigma(\alpha)}-\lambda_{\sigma(\beta)}}.
\label{rela2}
\end{equation}
It is easily checked that this is indeed the solution of all the
equations~(\ref{rela1}).

So, for \emph{any} set of values~$\{\lambda_\alpha\}$, the wave
function~(\ref{wavefunction}) where the~$\{a(\sigma)\}$ are given
by~(\ref{rela2}) is an eigenvector of the Hamiltonian~(\ref{hamilton}).
The values of the~$\{\lambda_\alpha\}$ can then be obtained from the
boundary conditions~(\ref{nbc}). One gets:
\begin{equation}
a(\sigma)=e^{z+\lambda_{\sigma(1)}}a(\sigma\circ{\cal C}),
\label{circ1}
\end{equation}
where~$\cal C$ is the circular permutation ${\cal C}(1)=2$, ${\cal
C}(2)=3$,\dots, ${\cal C}(n-1)=n$, 
${\cal C}(n)=1$.
Using~(\ref{rela2}), we easily get the new Bethe Ansatz equations.
For all~$\alpha$,
\begin{equation}
e^{\lambda_\alpha+z}=\prod_{\substack{1\le\beta\le
n\\\beta\ne\alpha}}{\lambda_\alpha-\lambda_\beta+\gamma
\over
\lambda_\alpha-\lambda_\beta-\gamma}.
\end{equation}

We are only interested in the ground-state solution. By continuity of
this ground state, we get the last condition
\begin{equation}
\lim_{\substack{z\to0\\\gamma\to0}}\lambda_\alpha=0.
\end{equation}

\section{Bethe Ansatz equations when all the~$\{z_i\}$ except~$z_1$ have the
same value~$z$} \label{z1zz}

When the parameters~$\{z_i\}$ do not take the same value~$z$,
the computation of the energy~$E$ is more complicated; indeed
the wave function~$\psi$ is no longer a symmetric function of
the~$\{x_i\}$ and there is no way that the standard Bethe
Ansatz~(\ref{wavefunction}) might lead to the result.

However, in order to  study the Hamiltonian~(\ref{hamilton}) for
fermionic particles or, more generally, for particles with arbitrary
symmetries and anti-symmetries, a more general Ansatz
than~(\ref{wavefunction}) has been proposed\cite{Yang.67, Sutherland.68,
Kardar.87, EmigKardar.01}: in~(\ref{wavefunction}) and~(\ref{deftau}),
the permutation~$\tau$ is only introduced as a convenient way to get the
coordinates~$\{x_i\}$ of the $n$ particles sorted from the leftmost
particle to the rightmost in the expression of the wave function. An easy
way to break the symmetry of~$\psi$ is to make the parameters
$\{a(\sigma)\}$ explicitly dependent on the permutation~$\tau$:
\begin{equation}
\psi_{z_1,\dots,z_n}(x_1,\dots,x_n)
=\sum_{\sigma}a(\tau,\sigma)e^{\sum_{i=1}^n\lambda_{\sigma(i)}x_{\tau(i)}},
\label{wavefunction2}
\end{equation}
where the permutation~$\tau$ is, as before, defined by~(\ref{deftau}).

As shown below, the solution to our problem with the unusual boundary
conditions~(\ref{nbc}) can also be written using~(\ref{wavefunction2}).
We first begin with the most general case where all the~$\{z_i\}$ are
different and, at some point, specialize to the simpler case where all
the~$\{z_i\}$ except $z_1$ are identical.

\subsection{General Bethe Ansatz equations for arbitrary~$\{z_i\}$.}

Using the Ansatz~(\ref{wavefunction2}) in~${\cal H}\psi=E\psi$
where~$\cal H$ the Hamiltonian~(\ref{hamilton}), we have, as usual:
\begin{equation}
E(n,\gamma;z_1,\dots, z_n)=-{1\over2}\sum_{\alpha=1}^n\lambda_\alpha^2.
\end{equation}
The new equations for the parameters~$\{a(\tau,\sigma)\}$ are more
complicated than~(\ref{rela1}):
\begin{equation}
a(\tau,\sigma\circ T_k) =
{\big(\lambda_{\sigma(k)}-\lambda_{\sigma(k+1)}\big)a(\tau\circ
T_k,\sigma)
- \gamma a(\tau,\sigma)\over
  \lambda_{\sigma(k)}-\lambda_{\sigma(k+1)}+\gamma},
\label{oneeq}
\end{equation}
for any permutations~$\tau$ and~$\sigma$ and for any integer~$1\le k<n$.

A convenient way to write the~$(n!)^2$ parameters~$\{a(\tau,\sigma)\}$ is
using~$n!$ vectors indexed by~$\sigma$, each vector having $n!$
components:
\begin{equation}
\vec{a(\sigma)}=\left|\begin{array}{l}
	a(\tau_1,\sigma)\\
	a(\tau_2,\sigma)\\
	\qquad\vdots\\
	a(\tau_{n!},\sigma)\end{array}\right.,
\end{equation}
where~$\tau_1$, \dots, $\tau_{n!}$ are the $n!$ permutations
of~$\{1,\dots,n\}$ sorted in an arbitrary way chosen once for all. (The
order must, of course, be the same for all values of~$\sigma$.) We now
introduce the matrices~$M_k$ defined by
\begin{equation}
\left|\begin{array}{l}
        a(\tau_1\circ T_k,\sigma)\\
        a(\tau_2\circ T_k,\sigma)\\
        \qquad\vdots\\
        a(\tau_{n!}\circ T_k,\sigma)\end{array}\right.
= M_k \vec{a(\sigma)}.
\end{equation}
Those matrices~$M_k$ just shuffle the components of the
vector~$\vec{a(\sigma)}$; there is thus exactly one ``1'' per raw and
per column and all the other components are ``0''. In a concise way, we
can write~$M_k$ as
\begin{equation}
\left(M_k\right)_{i,j}=\delta_{\tau_i}^{\tau_j\circ T_k}.
\end{equation}

The matrices~$M_k$ are a representation of the permutations~$T_k$. As
such, they have the same standard commutation properties as the
permutations:
\begin{equation}
\begin{gathered}
M_k^2=I,
\qquad M_k M_{k+1} M_k =M_{k+1} M_k M_{k+1},\\
M_k M_{k'}=M_{k'}M_k\qquad\text{if $|k-k'|>1$}.
\end{gathered}
\label{propm}
\end{equation}
($I$ being the identity matrix.)
Equation~(\ref{oneeq}) is then simply written as
\begin{equation}
\vec{a(\sigma\circ T_k)}=Y_k^{\sigma(k),\sigma(k+1)}\vec{a(\sigma)},
\label{rela3}
\end{equation}
where $Y_k^{i,j}$ is the Yang-Baxter operator defined\cite{Yang.67} by
\begin{equation}
Y_k^{i,j}={\big(\lambda_i-\lambda_j\big)M_k- \gamma I
	\over  \lambda_i-\lambda_j+\gamma}.
\label{Ydef}
\end{equation}
It is clear from~(\ref{rela3}) that any vector~$\vec{a(\sigma)}$ can be
obtained from the knowledge of one of them. However, as in the
symmetric case, one has to check that the result does not depend on the
way the permutations are decomposed as a product of the elementary
permutations~$T_k$. There are no explicit
formula\cite{LascouxLeclercThibon.97} such as~(\ref{rela2})
of~$\vec{a(\sigma)}$, but one can check that the~$(n-1)n!$
relations~(\ref{rela3}) are indeed self compatible. This is implied by
the following ``Yang Baxter'' relations\cite{Yang.67} 
\begin{equation}
\begin{gathered}
Y_k^{i,j} Y_k^{j,i} = I,\\
Y_k^{i,j} Y_{k'}^{i',j'} = Y_{k'}^{i',j'}Y_k^{i,j}\qquad\text{if
$|k-k'|>1$},\\
Y_k^{i,j} Y_{k+1}^{i,l} Y_k^{j,l} = Y_{k+1}^{j,l}Y_k^{i,l}Y_{k+1}^{i,j},
\end{gathered}
\label{Yprop}
\end{equation}
which can be easily checked using~(\ref{propm}) and (\ref{Ydef}).
With the first of those three relations, using~(\ref{rela3}) twice to
compute $\vec a(\sigma\circ T_k\circ T_k)$ gives correctly~$\vec
a(\sigma)$. The second relation implies $\vec a(\sigma\circ T_k\circ
T_{k'})=\vec a(\sigma\circ T_{k'}\circ T_k)$ if~$|k-k'|>1$ and, finally,
the third relation gives~$\vec a(\sigma\circ T_k\circ T_{k+1}\circ T_k) =
\vec a(\sigma\circ  T_{k+1}\circ T_k\circ T_{k+1})$. It is a well known
property of the symmetric group that those three necessary conditions
are actually sufficient to ensure that the relations~(\ref{rela3}) are
self consistent.

One still needs to write the boundary conditions~(\ref{nbc}) with the
parameters~$\{a(\tau,\sigma)\}$.
From~(\ref{wavefunction2}), one gets
\begin{equation}
a(\tau,\sigma)=
\exp({z_{\tau(1)}+\lambda_{\sigma(1)}})a(\tau\circ{\cal C},\sigma\circ{\cal C}).
\label{circ}
\end{equation}
where~${\cal C}$ is, as in~(\ref{circ1}), the circular permutation.

As~${\cal C}=T_1\circ T_2\circ\dots\circ T_{n-1}$, the matrix that
shuffles the lines of the vectors~$\vec{a(\sigma)}$ according to the
permutation~${\cal C}$ is simply the product of the matrices~$M_k$.
Thus, we have
\begin{equation}
\vec{a(\sigma)}=\exp({\lambda_{\sigma(1)}}) Z M_1 M_2\dots M_{n-1}
\vec{a(\sigma\circ {\cal C})},
\label{circv}
\end{equation}
where~$Z$ is the diagonal matrix defined by
\begin{equation}
(Z)_{i,j}=\delta_i^j \exp({z_{\tau_i(1)}}).
\label{defZ}
\end{equation}
Moreover, using several times~(\ref{rela3}), we get, from the definition
of~${\cal C}$,
\begin{equation}
\vec a(\sigma\circ{\cal C})=Y_{n-1}^{\sigma(1),\sigma(n)}
Y_{n-2}^{\sigma(1),\sigma(n-1)}\dots Y_1^{\sigma(1),\sigma(2)}\vec
a(\sigma).
\label{circv2}
\end{equation}

Putting together~(\ref{circv}) and~(\ref{circv2}), we see that $\vec
a(\sigma)$ must be, for each~$\sigma$, the eigenvector of some operator.
\begin{equation}
\exp(\lambda_{\sigma(1)})Z M_1\dots M_{n-1} Y_{n-1}^{\sigma(1),\sigma(n)}
\dots Y_1^{\sigma(1),\sigma(2)}\vec
a(\sigma)= \vec a(\sigma).
\label{operator}
\end{equation}
There exists a non-zero~$\vec a(\sigma)$ such as~(\ref{operator}) holds
only for certain values of the~$\{\lambda_\alpha\}$.
However, as the $n!$ vectors $\{\vec a(\sigma)\}$ are not independent
variables, we need to check that the $n!$ relations~(\ref{operator}) are
compatible: they must hold
simultaneously for the same values of the~$\{\lambda_\alpha\}$.
As explained in appendix~\ref{compatible}, this is the
case. To obtain the values of the~$\{\lambda_\alpha\}$ we are looking
for, we use~(\ref{operator}) when~$\sigma$ is the identical permutation.
A non-zero~$\vec a(\sigma)$ exists if and only if
\begin{equation}
\det\left(I-\exp(\lambda_1)Z M_1\dots M_{n-1}Y_{n-1}^{1,n}
Y_{n-2}^{1,n-1}\dots Y_1^{1,2}\right)=0,
\end{equation}
or, using the properties~(\ref{Yprop}) of the operators~$Y$,
\begin{equation}
\det\left(Y_1^{2,1}Y_2^{3,1}\dots Y_{n-1}^{n,1}-\exp(\lambda_1)Z M_1\dots
M_{n-1}\right)=0,
\end{equation}
or, using the definition~(\ref{Ydef}) of the operators~$Y$,
\begin{equation}
\det\left(\prod_{\alpha=1}^{n-1}{(\lambda_1-\lambda_{\alpha+1})M_\alpha+\gamma I\over
\lambda_1+\lambda_{\alpha+1}-\gamma}-\exp(\lambda_1)Z\prod_{\alpha=1}^{n-1}M_\alpha
\right)=0.
\label{bigdet}
\end{equation}
That last equation relates~$\exp(\lambda_1)$ to the~$\{\lambda_\alpha\}$.
There are~$n-1$ other equations giving all the~$\exp(\lambda_k)$ which we
can obtain either by using~(\ref{operator}) with different
permutations~$\sigma$ either, as they play symmetric roles, by shuffling
the~$\{\lambda_\alpha\}$ in~(\ref{bigdet}).

Finally, the wave function~(\ref{wavefunction2}) introduced is indeed an
eigenvector of the Hamiltonian~(\ref{hamilton}) with the boundary
conditions~(\ref{nbc}), provided that the~$\{\lambda_\alpha\}$ are such
that~(\ref{bigdet}) and the~$n-1$ other relations obtained by symmetry
hold.

Note that the new Bethe Ansatz equation~(\ref{bigdet}) can be regarded as
a polynomial of degree~$n!$ in~$\exp(\lambda_1)$, so that we have not one
value of~$\exp(\lambda_1)$ as a function of
the~$\{\lambda_\alpha\}$, but~$n!$. This could be expected, as the
method we
have used is known to generate, when~$z_i=0$, not only the bosonic
solution, but all the eigenvalues of~(\ref{hamilton}) for arbitrary
symmetry relations between the particles. So, if we write
from~(\ref{bigdet}) the~$n!$ possible expressions of~$\exp(\lambda_1)$
and make the~$\{z_i\}$ go to zero, we will recover the usual bosonic
Bethe Ansatz solution~(\ref{e2}), but also the fermionic
solution~$\exp(\lambda_\alpha)=1$ and all the intermediate cases. In our
problem, we are looking for the ground state energy of discernible
particles and it is known, in this situation, that the ground state
is given by the bosonic solution.

To sum up, what remains to be done is to single out from~(\ref{bigdet})
the expression of~$\exp(\lambda_1)$ which goes to the standard bosonic
equations~(\ref{e2}) when the~$\{z_i\}$ vanish, to write by symmetry
the~$n-1$ remaining equations giving all the~$\{\exp(\lambda_\alpha)\}$
as functions of the~$\{\lambda_\alpha\}$, to solve those non-algebraic
equations in order to write the ground state
energy~$E=-(1/2)\sum\lambda_\alpha^2$, and, finally, to take the
limit~$n\to0$ and various derivatives with respect to the~$z_i$ to obtain
the different quantities characterizing the winding number of the
polymer.

As this seems to be a difficult task in the general case, we will go
through this program only when all the~$\{z_i\}$ have the same value~$z$
except for~$z_1$.

\subsection{Simplification when all the~$\{z_i\}$ are equal except for~$z_1$.}

When \emph{all} the~$\{z_i\}$ are set to zero, the matrix~$Z$ is the
identity matrix and one of the solutions of~(\ref{bigdet}) must be the
standard bosonic Bethe Ansatz equation~(\ref{e2}).
One way to see it is to notice that, when~(\ref{e2}) holds, the
vector which cancels the matrix
in~(\ref{bigdet}) is simply~$(1,\dots,1)$.
Another way to see it is to notice that to derive~(\ref{bigdet}), we
never actually used the matrix representation of~$M_k$, $Y_k^{i,j}$,
etc., but only the commutation properties of those matrices. If we were
to choose other \emph{representations} of those matrices having the same
commutation properties, relation~(\ref{bigdet}) would still be valid with
those representations. For example, when all the~$\{z_i\}$ are zero, the
bosonic solution is obtained from~(\ref{bigdet}) by choosing the trivial
representation~$M_k=1$. The fermionic solution~$\exp(\lambda_\alpha)=1$ is
obtained by choosing~$M_k=-1$, etc.

In the situation where~$z_i=z$ for~$i\ge2$, with only~$z_1$ different
from~$z$, we can make a similar simplification. Indeed, in that case, the
particles~$x_2$, \dots, $x_n$ play symmetric roles. Thus, the ground
state solution must be symmetric in those variables. Back to the wave
function~(\ref{wavefunction2}), this means that the
parameters~$a(\tau_1,\sigma)$ and~$a(\tau_2,\sigma)$ must be equal
if~$\tau_1^{-1}(1)=\tau_2^{-1}(1)$. In other words, the
parameters~$a(\tau,\sigma)$
do not depend on the whole shuffling~$\tau$ of the particles~$\{x_i\}$,
but only on the position of~$x_1$ relatively to the other; for
each~$\sigma$, there is one value of~$a(\tau,\sigma)$ corresponding to
the first particle being the leftmost, another value when the first
particle is the second leftmost, etc.

This suggests that equation~(\ref{bigdet}) can be written in that
situation with a representation of the~$M_k$ as matrices of size~$n\times
n$ instead of~$n!\times n!$. Indeed, we write the new vector~$\vec
a(\sigma)$ as
\begin{equation}
\vec a(\sigma)=\left|\begin{array}{l}
a(\tau,\sigma)\text{ such that~$\tau(1)=1$ ($x_1$ is the leftmost
particle)}\\
a(\tau,\sigma)\text{ such that~$\tau(2)=1$ ($x_1$ is the second leftmost
particle)}\\
\qquad\vdots\\
a(\tau,\sigma)\text{ such that~$\tau(n)=1$ ($x_1$ is the rightmost
particle)}
\end{array}\right.
\end{equation}
Then, the matrix~$M_k$ that switches the~$k$-th and~$k+1$-th particles
is given by
\begin{equation}
(M_k)_{i,j}=\begin{cases}
	\delta_j^{k+1}&\text{if $i=k$},\\
	\delta_j^k&\text{if $i=k+1$},\\
	\delta_i^j&\text{otherwise},
\end{cases}\end{equation}
that is:
\begin{equation}
M_1=\left(\begin{matrix}
	\textbf{0}&\textbf{1}&0&0&\dots\\
	\textbf{1}&\textbf{0}&0&0&\dots\\
	0&0&1&0&\dots\\
	0&0&0&1&\dots\\
	\hdotsfor{4}&\ddots\end{matrix}\right),
\qquad
M_2=\left(\begin{matrix}
	1&0&0&0&\dots\\
	0&\textbf{0}&\textbf{1}&0&\dots\\
	0&\textbf{1}&\textbf{0}&0&\dots\\
	0&0&0&1&\dots\\
	\hdotsfor{4}&\ddots\end{matrix}\right),
\end{equation}
etc. With all the~$\{z_i\}$ but~$z_1$ equal to~$z$, we can also write the
new matrix~$Z$ in this representation. It is a diagonal matrix, on the first
line we have~$\exp(z_1)$ as $x_1$ is then the leftmost particle, and, on the
other lines, we do not know which particle is the leftmost, but it is of
no importance as we know it is not~$x_1$ and as all the other particles have the
same parameter~$z$. Thus, we have:
\begin{equation}
Z=\left(\begin{matrix}
	e^{z_1}&0&0&\dots\\
	0&e^z&0&\dots\\
	0&0&e^z&\dots\\
	\hdotsfor{3}&\ddots\end{matrix}\right).
\end{equation}

\emph{Of course}, the new matrices~$M_k$ have the correct commutation
relations~(\ref{propm}) and the final result~(\ref{bigdet}) is still
valid with the new matrices~$M_k$ and~$Z$.

\subsection{Explicit expression of the determinant}

When~$z_i=z$ for $i\ge2$, using the new matrices~$M_k$ and~$Z$,
we can compute the determinant in~(\ref{bigdet}) by induction.
First, we normalize~$\exp(\lambda_1)$ by the standard Bethe Ansatz
expression and write
\begin{equation}
e^{\lambda_1+\zeta_1}=\prod_{\alpha=2}^n{\lambda_1-\lambda_\alpha+\gamma\over
	\lambda_1-\lambda_\alpha-\gamma},
\end{equation}
where~$\zeta_1$ is the quantity we are trying to determine.

Using this new variable, we write the determinant in~(\ref{bigdet}), up
to a multiplicative prefactor, as
\begin{equation}
d_n=\det\big(A_n-\exp(-\zeta_1) B_n\big),
\end{equation}
where
\begin{equation}
A_n=\prod_{\alpha=1}^{n-1}{(\lambda_1-\lambda_{\alpha+1})M_\alpha+\gamma\over
\lambda_1-\lambda_{\alpha+1}+\gamma}
\label{An}
\end{equation}
and, as can easily be seen,
\begin{equation}
B_n=Z\prod_{\alpha=1}^{n-1}M_\alpha=\left(\begin{matrix}
	0	& 0	& \dots	& 0	& e^{z_1}	\\
	e^z	& 0	& \dots	& 0	& 0		\\
	0	& e^z	& \dots	& 0	& 0		\\
	\hdotsfor{2}&\ddots&\hdotsfor{2}\\
	0	& 0	& \dots & e^z	& 0
	\end{matrix}\right).
\end{equation}
As~$e^{z_1}$ appears only once in the matrix, the determinant~$d_n$
can be written as
\begin{equation}
d_n=\alpha_n-e^{z_1}\beta_n,
\label{alphabeta}
\end{equation}
where~$\alpha_n$ and $\beta_n$ do not depend on~$z_1$.

Going from~$A_n$ to $A_{n+1}$ is easy enough; singling out the last term
in the product~(\ref{An}), one has
\begin{equation}
A_{n+1}=\left(\begin{matrix}
\boxed{\begin{matrix}\\&A_n&\\ \\\end{matrix}}&
\begin{matrix}\vdots\\0\\\vdots\end{matrix}\\
\dots\ 0\ \dots&1
\end{matrix}\right)\cdot\left(\begin{matrix}
	1\\
	 &\ddots&&{\ \ (0)}\\
	 && 1 \\
	 &{(0)}&  &\strut \mu_{n+1} & 1-\mu_{n+1}\\
	 &&  & 1-\mu_{n+1} & \mu_{n+1}\end{matrix}\right),
\end{equation}
with \begin{equation}
\mu_{n+1}={\gamma\over\lambda_1-\lambda_{n+1}+\gamma}.\end{equation}
Doing this last multiplication, we see that
the first~$n-1$ columns of~$A_{n+1}$ are the first columns of~$A_n$
padded with one final zero, and that the~$n$-th and~$n+1$-th columns
of~$A_{n+1}$ are the~$n$-th column of~$A_n$ with different
multiplicative factors (respectively~$\mu_{n+1}$ and~$1-\mu_{n+1}$) and different
paddings (respectively $1-\mu_{n+1}$ and~$\mu_{n+1}$.) Thus, if we develop~$d_{n+1}$
over the last line, there are only two terms which look very much
like~$d_n$. The only differences are that the last column is multiplied
by some factor and that the term~$e^{z_1}$ is either missing or not
multiplied by the numerical factor that affects its column. Finally,
using~(\ref{alphabeta}), one can get
\begin{equation}
d_{n+1}=\mu_{n+1}\big(\mu_{n+1}\alpha_n\big)-\big(1-\mu_{n+1}-e^{z-\zeta_1}\big)\big((1-\mu_{n+1})\alpha_n
-\beta_ne^{z_1}\big),
\end{equation}
or
\begin{equation}
\alpha_{n+1}=\big(2\mu_{n+1}-1+(1-\mu_{n+1})e^{z-\zeta_1}\big)\alpha_n,\quad
\beta_{n+1}=-\big(1-\mu_{n+1}-e^{z-\zeta_1}\big)\beta_n
\end{equation}
with~$\alpha_1=1$ and $\beta_1=\exp(-\zeta_1).$
It is now easy to compute~$\alpha_n$ and~$\beta_n$:
\begin{equation}
\begin{gathered}
\alpha_n=\prod_{i=2}^n{(\lambda_1-\lambda_i)(e^{z-\zeta_1}-1)+\gamma
	\over \lambda_1-\lambda_i+\gamma},\\
\beta_n=e^{-\zeta_1}\prod_{i=2}^n{(\lambda_1-\lambda_i)(e^{z-\zeta_1}-1)
		+\gamma e^{z-\zeta_1}
	\over \lambda_1-\lambda_i+\gamma},
\end{gathered}\end{equation}
and, finally, the
condition that the determinant~(\ref{bigdet}) is zero gives the
following
result:
\begin{equation}
e^{z}\prod_{i=1}^n\left[1+{\lambda_1-\lambda_i\over\gamma}\left(e^{z-\zeta_1}-1\right)\right]=e^{z_1}\prod_{i=1}^n\left[e^{z-\zeta_1}+{\lambda_1-\lambda_i\over\gamma}\left(e^{z-\zeta_1}-1\right)\right].
\end{equation}
As all the~$\{\lambda_\alpha\}$ play symmetric roles, this is exactly the
result announced~(\ref{2prod}).

\section{Proof equation~(\ref{operator}) can be satisfied simultaneously
for all permutations~$\sigma$.}
\label{compatible}

To prove that equations~(\ref{operator}) are indeed compatible, we start by
assuming that the~$\{\vec a(\sigma)\}$ are such that~(\ref{rela3})
holds for any~$\sigma$ and~$k$.  As a consequence, (\ref{circv2}) is
true for any permutation and~(\ref{operator}) is equivalent
to~(\ref{circv}).

Furthermore, we assume that~(\ref{operator}) (or (\ref{circv})) is true
for a given permutation~$\sigma$. To show that it is also true for any
other permutation, it is sufficient, by induction, to prove
that~(\ref{operator}) (or (\ref{circv})) holds for~$\sigma\circ T_k$
with~$1\le k<n$.

It is necessary to distinguish the two cases $k\ne1$ and $k=1$:

\subsubsection*{$\bullet$ When~$1<k<n$}

When~$k\ne1$, we have the following properties:
\begin{equation}
\begin{gathered}
T_k\circ{\cal C}={\cal C}\circ T_{k-1},\qquad Y_k^{i,j}Z=Z Y_k^{i,j},\\
Y_k^{i,j} M_1 M_2\dots M_{n-1}= M_1 M_2\dots M_{n-1}
Y_{k-1}^{i,j}.
\end{gathered}
\label{kne1}
\end{equation}
The first relation is a basic property of permutations, the second
relation comes from the fact that~$Y_k^{i,j}$ does not change
the value of~$\tau(1)$, and the third one, considering the
definition~(\ref{Ydef}) of~$Y_k$, is a rewriting of the first relation
in the matrix representation.

We can now show that~(\ref{circv}) and, therefore, (\ref{operator})
holds for $\sigma\circ T_k$:
\begin{equation}
\begin{aligned}
\vec a(\sigma\circ T_k) &=
Y_k^{\sigma(k),\sigma(k+1)}\vec a(\sigma)\\
&=Y_k^{\sigma(k),\sigma(k+1)}\exp({\lambda_{\sigma(1)}}) Z M_1 M_2\dots M_{n-1}
\vec{a(\sigma\circ {\cal C})},\\
&=\exp({\lambda_{\sigma(1)}}) Z M_1 M_2\dots
M_{n-1}Y_{k-1}^{\sigma(k),\sigma(k+1)}\vec{a(\sigma\circ {\cal C})},\\
&=\exp({\lambda_{\sigma(1)}}) Z M_1 M_2\dots
M_{n-1}\vec a(\sigma\circ{\cal C}\circ T_{k-1})\\
&=\exp({\lambda_{\sigma(1)}}) Z M_1 M_2\dots
M_{n-1}\vec a(\sigma\circ T_k\circ{\cal C}).
\end{aligned}
\end{equation}
As~$\sigma(1)=\sigma\circ T_k(1)$ for~$k\ne1$, this is indeed~(\ref{circv})
applied to the permutation~$\sigma\circ T_k$.

\subsubsection*{$\bullet$ When~$k=1$}

Equation~(\ref{operator}) express that~$\vec a(\sigma)$ is an eigenvector of
\begin{equation}
A=ZM_1\dots M_{n-1}
Y_{n-1}^{\sigma(1),\sigma(n)}
Y_{n-2}^{\sigma(1),\sigma(n-1)}\dots Y_2^{\sigma(1),\sigma(3)}Y_1^{\sigma(1),\sigma(2)},
\end{equation}
and we want to prove that~$\vec a(\sigma\circ T_1)$ is also an eigenvector of
\begin{equation}
ZM_1\dots M_{n-1}
Y_{n-1}^{\sigma(2),\sigma(n)}
Y_{n-2}^{\sigma(2),\sigma(n-1)}\dots
Y_2^{\sigma(2),\sigma(3)}Y_1^{\sigma(2),\sigma(1)}.
\end{equation}
As~$\vec a(\sigma\circ T_1)=Y_1^{\sigma(1),\sigma(2)}\vec
a(\sigma)$, this is equivalent to prove that~$\vec a(\sigma)$ is an
eigenvector of
\begin{equation}
B= Y_1^{\sigma(2),\sigma(1)}ZM_1\dots M_{n-1}
Y_{n-1}^{\sigma(2),\sigma(n)}
Y_{n-2}^{\sigma(2),\sigma(n-1)}\dots Y_2^{\sigma(2),\sigma(3)}.
\end{equation}
(The relation $Y_k^{i,j}Y_k^{j,i}=I$ has been used twice.)

To conclude, we presently show that~$AB=BA$, which implies that $A$ et $B$
have the same eigenvectors.

First, we define another diagonal matrix~$Z_2$ by
\begin{equation}
\left(Z_2\right)_{i,j}=\delta_i^j \exp(z_{\tau_i(2)}).
\end{equation}
(Compare with~(\ref{defZ}).) Clearly, we have
\begin{equation}
M_1 Z = Z_2 M_1\qquad\text{and}\qquad M_1 Z_2 = Z M_1.
\end{equation}
Moreover, as a consequence, the product~$Z Z_2=Z_2 Z$ commutes with $M_1$
and~$Y_1$.

When computing~$AB$, two matrices $Y_1$ cancel out. The matrix~$Z$ commutes
with all the the~$Y_k$ and all the~$M_k$ except $Y_1$ and $M_1$, so that
we can ``move'' the second~$Z$  to the left and obtain
\begin{equation}
\begin{aligned}
AB &= Z (\prod M_i) Y_{n-1}^{\sigma(1),\sigma(n)}\!\dots
Y_2^{\sigma(1),\sigma(3)} Z (\prod M_i)
Y_{n-1}^{\sigma(2),\sigma(n)}\!\dots Y_{2}^{\sigma(2),\sigma(3)},\\
&= Z Z_2 (\prod M_i) Y_{n-1}^{\sigma(1),\sigma(n)}\!\dots
Y_2^{\sigma(1),\sigma(3)} (\prod M_i)
Y_{n-1}^{\sigma(2),\sigma(n)}\!\dots Y_{2}^{\sigma(2),\sigma(3)}.
\end{aligned}\end{equation}
When computing~$BA$, the second matrix~$Z$
can also travel to the left; we get:
\begin{equation} \begin{aligned}
BA&=Y_1^{\sigma(2),\sigma(1)}Z(\prod M_i)
Y_{n-1}^{\sigma(2),\sigma(n)}\dots Y_2^{\sigma(2),\sigma(3)}
Z(\prod M_i)\\
&\hphantom{=}\times Y_{n-1}^{\sigma(1),\sigma(n)}\dots
Y_1^{\sigma(1),\sigma(2)}\\
&=Y_1^{\sigma(2),\sigma(1)}Z Z_2 (\prod M_i)
Y_{n-1}^{\sigma(2),\sigma(n)}\dots Y_2^{\sigma(2),\sigma(3)}
(\prod M_i) Y_{n-1}\dots\\
&=Z Z_2 Y_1^{\sigma(2),\sigma(1)}(\prod M_i)
Y_{n-1}^{\sigma(2),\sigma(n)}\dots Y_2^{\sigma(2),\sigma(3)}
(\prod M_i) Y_{n-1}\dots
\end{aligned} \end{equation}
Thus, the products $AB$ and $BA$ share the same prefactor~$ZZ_2$, so
that if $AB=BA$ is true when~$z_i=0$ (the case studied by
Yang\cite{Yang.67}), then~$AB=BA$ is true for arbitrary values of
the~$\{z_i\}$. As it is a well known fact that the operators commute
in Yang's case, we could stop the proof here. However, for completeness,
let us properly finish it.

We continue the simplification of~$AB$; using~(\ref{kne1}), we can have
the whole first group of matrices~$Y_k$ in~$AB$ go to the right
through the second product~$M_1\dots M_{n-1}$. We get
\begin{equation}
AB=ZZ_2(\prod M_i)^2
Y_{n-2}^{\sigma(1),\sigma(n)}\dots Y_1^{\sigma(1),\sigma(3)}
Y_{n-1}^{\sigma(2),\sigma(n)}\dots Y_2^{\sigma(2),\sigma(3)}.
\end{equation}
We do the same for the product~$BA$:
\begin{equation}
BA=Z Z_2 Y_1^{\sigma(2),\sigma(1)}(\prod M_i)^2
Y_{n-2}^{\sigma(2),\sigma(n)}\dots Y_1^{\sigma(2),\sigma(3)}
Y_{n-1}^{\sigma(1),\sigma(n)}
\dots Y_1^{\sigma(1),\sigma(2)}.
\end{equation}
Using 
\begin{equation}
Y_1^{i,j}(M_1\dots M_{n-1})^2=(M_1\dots M_{n-1})^2Y_{n-1}^{i,j},
\end{equation}
which can be deduced from the properties~(\ref{propm})
of the matrices~$M_k$, we get
\begin{equation}
BA=ZZ_2(\prod M_i)^2
Y_{n-1}^{\sigma(2),\sigma(1)}Y_{n-2}^{\sigma(2),\sigma(n)}\dots
Y_1^{\sigma(2),\sigma(3)}
 Y_{n-1}^{\sigma(1),\sigma(n)}
\dots Y_1^{\sigma(1),\sigma(2)}.
\end{equation}
$AB$ and $BA$ have the same prefactor$ZZ_2(\prod M_i)^2$; we need to show
that the two products of matrices~$Y_k$ are equal. We proceed by
induction: It is clear for~$n=1$ (or $n=2$) and we assume it is true
for~$n-1$. In both products, we ``move'' the matrices~$Y_{n-1}$ to the
left. We get:
\begin{equation}\begin{aligned}
AB&=ZZ_2(\prod M_i)^2
Y_{n-2}^{\sigma(1),\sigma(n)}Y_{n-1}^{\sigma(2),\sigma(n)}
 Y_{n-3}^{\sigma(1),\sigma(n-1)}\dots
Y_1^{\sigma(1),\sigma(3)}\\
&\hphantom{=}\times
Y_{n-2}^{\sigma(2),\sigma(n-1)}\dots Y_2^{\sigma(2),\sigma(3)}
\end{aligned}\end{equation}
and, using~(\ref{Yprop}),
\begin{equation}
\begin{aligned}
BA&=ZZ_2(\prod M_i)^2 Y_{n-1}^{\sigma(2),\sigma(1)}Y_{n-2}^{\sigma(2),\sigma(n)}
Y_{n-1}^{\sigma(1),\sigma(n)} Y_{n-3}^{\sigma(2),\sigma(n-1)}\dots
Y_1^{\sigma(2),\sigma(3)}\\
&\hphantom{=}\times Y_{n-2}^{\sigma(1),\sigma(n-1)}
\dots Y_1^{\sigma(1),\sigma(2)},\\
&=ZZ_2(\prod M_i)^2 Y_{n-2}^{\sigma(1),\sigma(n)}
Y_{n-1}^{\sigma(2),\sigma(n)}Y_{n-2}^{\sigma(2),\sigma(1)}
Y_{n-3}^{\sigma(2),\sigma(n-1)}\dots Y_1^{\sigma(2),\sigma(3)}\\
&\hphantom{=}\times Y_{n-2}^{\sigma(1),\sigma(n-1)}
\dots Y_1^{\sigma(1),\sigma(2)}.
\end{aligned}
\end{equation}
Leaving aside the common prefix $ZZ_2(\prod M_i)^2$, the products
$AB$ and $BA$ start with the same two~$Y$ matrices, and what remains are
the products of~$Y$ matrices in the expressions of~$AB$ and~$BA$ at
order~$n-1$.  This, by induction, proves that~$AB=BA$.

\bigbreak

Finally, putting everything together, we have shown that the~$n!$
properties~(\ref{circv}) obtained from the boundary conditions are self
compatible.

\end{document}